\documentstyle[prl,aps]{revtex}
\begin{document}
\input{epsf}
\draft
\twocolumn[\hsize\textwidth\columnwidth\hsize\csname@twocolumnfalse\endcsname
\title{Fluctuations provide strong selection in
Ostwald ripening}
\author{Baruch Meerson \cite{adr}}
\address{The Racah Institute  of  Physics, Hebrew
University   of  Jerusalem,
Jerusalem 91904, Israel}
\maketitle
\begin{abstract}
A selection problem that appears 
in the Lifshitz-Slyozov (LS) theory of Ostwald ripening is reexamined. 
The problem concerns selection of a self-similar
distribution function (DF) of the minority domains with respect to their 
sizes from a whole one-parameter
family of solutions. A strong
selection rule is found via an
account of
fluctuations. Fluctuations produce an infinite
tail in the DF and drive the DF towards
the ``limiting solution" of LS or its analogs for other growth mechanisms. 
\end{abstract}
\vskip1pc]
\narrowtext

Ostwald ripening (OR) \cite{Ostwald} is a fascinating and
generic process of self-organization in a physical
system far of equilibrium. It 
develops 
in a late stage of a first-order phase transition, in two or three
dimensions,
when a two-phase mixture undergoes coarsening and
the interfacial energy decreases subject to a global
conservation law \cite{LS,W}. OR continues to attract considerable 
attention
both in experiment \cite{experiment} and in 
theory \cite{theory,Brown,Chen,MS96,GMS}. For a nonlinear physicist, 
the problem of OR is of great interest because of a long-standing
selection problem \cite{LS,Brown,Chen,MS96,GMS} addressed below. 

Lifshitz and Slyozov (LS) \cite{LS} and Wagner \cite{W} developed a
mean-field formulation of OR, valid in the limit of a 
negligibly small volume fraction of the 
minority domains. In this formulation, the dynamics  
of the distribution function (DF) 
$F(R,t)$  of the minority domain sizes 
are governed (in scaled variables) by a continuity
equation,
\begin{equation} 
\frac{\partial F}{\partial t}+\frac{\partial}{\partial R}
(V F)=0\,,\quad 
V (R,t)= \frac{1}{R^n} \left(\frac{1}{R_c}-\frac{1}{R}\right)\,,
\label{1}
\end{equation}
where $R_c (t)$ is the critical radius for expansion/shrinkage of 
an individual
domain, while $n$ 
is determined by the growth mechanism. (For a review of different growth
mechanisms see Refs. \onlinecite{SS}. Most known are diffusion-controlled
growth, $n=1$, and interface-controlled growth, $n=0$.)
The dynamics
are constrained by conservation of the total mass (or volume) of the minority 
domains:
\begin{equation} 
\int_{0}^{\infty} R^3\,F(R,t)\,dR \,= Q = const \,.
\label{2}
\end{equation}

Scaling analysis of 
Eqs. (\ref{1}) and (\ref{2}) yields a similarity 
ansatz $F(R,t)=t^{-4/z}\, \Phi \,(R\,t^{-1/z})$ and $R_c= (t/\sigma)^{1/z}$,
where
$z = n+2\,$ and $\sigma=const$. Upon substitution, one obtains
a {\it family} of self-similar DFs
(formally, for every $n \ge -1$). Each of
the DFs is localized on a finite interval 
$[0, u_m]$
of the similarity variable 
$u=R\,t^{-1/z}$. The self-similar DFs can be parameterized by
$\sigma$, and there is a finite 
interval of allowed values of the scaled coefficient
$\sigma$. For each solution, the
average domain radius and the critical radius grow 
in time like $t^{1/z}$, while the concentration  
of domains decreases like $t^{-3/z}$. However, the {\it coefficients} in 
these scaling laws are $\sigma$-dependent. The scaling function $\Phi (\xi)$ 
has a 
markedly different shape depending on the value of $\sigma$. It should 
be stressed that the problem of OR, as described by Eqs. 
(\ref{1}) and (\ref{2}) is fully determined (of course, if one
prescribes an initial condition). Therefore, the scaled coefficient $\sigma$
is 
an observable quantity. It can be determined in a direct 
experiment or simulation 
by measuring, at large times,
the {\it coefficient} in the power law for the critical radius $R_c$ 
versus time.  The reader is referred to Ref. \onlinecite{GMS}
for a detailed description of the family of self-similar DFs for different 
values of $n$.

An important
selection problem therefore 
arises. It has a long history \cite{LS,Brown,Chen,MS96,GMS}, 
and its present status is as follows. There is only a ``weak" selection  
within the framework of the
``classical" model (\ref{1}) and (\ref{2}). The ``weak" selection rule, 
obtained
recently \cite{MS96,GMS}, is
the following. If the initial
DF $F(R,0)$ has a compact support $(0,R_m)$ and is 
describable by a power law $A (R_m-R)^{\lambda}$
in the close vicinity of $R=R_m$, then it is the exponent 
$\lambda$ that selects the correct self-similar asymptotic DF. The selected
value of parameter $\sigma$ is
\begin{equation}
\sigma = \frac{v_0^{n+2}}{(n+2) (v_0-1)}\,,
\label{a}
\end{equation}
where 
\begin{equation}
v_0= \frac{(n+2) \lambda + n + 5}{(n+1) \lambda + n + 4}
\label{b}
\end{equation}
(see Ref. \onlinecite{GMS}
for details). 
The celebrated
``limiting"
DFs obtained by LS \cite{LS} for $n=1$, and by
Wagner \cite{W} for $n=0$, correspond to 
{\it extended} 
initial DF or, formally, to $\lambda \to +\infty$. In this case Eqs. (\ref{a}) 
and (\ref{b}) yield the well-known ''universal" value 
$\sigma = 9/4$ for the diffusion-controlled OR, $n=1$, found by LS \cite{LS}.
On the contrary, as it is clear from Eqs. (\ref{a}) and (\ref{b}), for 
initial DFs with 
compact support and finite $\lambda$, different 
values of the scaled coefficient
$\sigma$ are obtained. Finally, 
for those 
initial DFs with compact support
that cannot be described by a power-law asymptotics
near $R=R_m$, convergence to {\it any}
self-similar solution is impossible \cite{Pego}.
A rigorous mathematical proof of these results 
is presently available \cite{Pego}. 

It has become clear after the analyses of Refs. \onlinecite{MS96,GMS,Pego} 
that
in order to get strong selection, one must go beyond the ``classical" model.
In the present communication I report on a 
progress in this direction. Here is an outline. I will
employ a mean-field cluster formulation of 
the problem
and proceed to the long-time limit, when only large clusters and single atoms
dominate. Using the characteristic 
inverse number of atoms in a cluster as 
a small parameter, I will arrive 
at a Fokker-Planck (FP) equation for
the cluster size distribution function. The drift 
term of the FP equation describes growth/shrinkage of clusters
due to an interplay between 
attachments and detachments of single atoms, and it
corresponds to the ``classical" LS-theory. 
The diffusion term of the FP equation accounts for
fluctuations, and it
is not present in the LS-theory. This term
becomes irrelevant at long times, however, it can play a very important role.
Indeed, even if the initial DF
has compact support, the diffusion term produces an infinite tail in the DF. 
As the result, the
DF will finally approach the limiting solution of LS \cite{LS} (or
its analogs for other growth mechanisms).

The mean-field rate 
equations of the cluster model
(see, {\it e.g.}, \cite{Binder,Gunton,Vvedensky}) represent a natural
extension of the mean-field 
continuum models, as these equations account for the discrete nature of
atoms: 
\begin{equation}
\dot{N_1} = -2 K_1 N_1^2 - N_1 \sum_{s \geq 2}K_s N_s +
2\frac{N_2}{\tau_2}+\sum_{s\geq3}\frac{N_s}{\tau_s}\,,
\label{3}
\end{equation}
\begin{equation}
\dot{N_s} = N_1 (K_{s-1} N_{s-1} - K_s N_s) -\frac{N_s}{\tau_s}
+\frac{N_{s+1}}{\tau_{s+1}}\,.
\label{4}
\end{equation}
Here $N_s (t)$ is the cluster size distribution function
($s$ is the number of atoms in a cluster), the $K_s$
are the rates of attachment of single atoms to a cluster of 
size $s$, and
the $\tau_s$ are the inverse rates of
detachment of single atoms from a cluster of size $s$. Rare events of
direct inter-cluster
coalescence (coagulation) are neglected in Eq. (\ref{4}); this
requires a small volume fraction of the ``cluster phase".
As no
external source of atoms is present in Eqs. (\ref{3})
and (\ref{4}), these equations preserve the total concentration of atoms: 
\begin{equation}
\frac{d}{dt}\sum_{s=1}^{\infty} s N_s (t) = 0\,,
\label{5}
\end{equation}
a discrete equivalent of Eq. (\ref{2}).

For most growth mechanisms (including the growth processes, controlled
by diffusion and by interface), 
the attachment-detachment kinetics, combined with 
mass conservation, promotes growth of larger
clusters at the expense of smaller ones, and this process 
is nothing but OR. Therefore, if 
the total concentration of atoms is large enough, the system
undergoes coarsening: the average cluster size grows in time, 
and the total
number
of clusters decreases. The late time asymptotics of this process
should reproduce 
OR quantitatively \cite{Binder,Gunton}. At the coarsening stage 
the number of
clusters with small $s$ become very small, except 
for the concentration of single atoms
$N_1$ that remains relatively large because of the ongoing 
detachment processes.
Therefore, one should consider the population of single atoms separately. 
As far as descriptions of clusters is concerned, one can
 proceed to the
limit of $s \gg 1$, treat $s$ as a continuous variable and use
Taylor expansion in $1/s$ in Eqs. (\ref{3}) and (\ref{4}). Essentially, this 
derivation follows the paper of
Binder \cite{Binder}. However, 
in order to account for fluctuations, we should
keep the second 
order terms in $1/s$ (in contrast to the approach of Binder who
kept such terms in his description of
the nucleation stage,
but neglected them in the coarsening stage). As the result, 
Eq. (\ref{4}) takes the form of a FP equation:
\begin{equation}
\frac{\partial N_s}{\partial t} + \frac{\partial}{\partial s}(V_s N_s)
=\frac{1}{2}\, \frac{\partial ^2}{\partial s^2} (D_s N_s)\,,
\label{6}
\end{equation}
where 
\begin{equation}
V_s (t) = K_s N_1 -1/\tau_s \quad \mbox{and} \quad D_s (t) = 
K_s N_1 + 1/\tau_s
\label{6a}
\end{equation}
are the drift velocity 
and diffusion coefficient
in the $s$-space. 

A continuous version of equation for $\dot{N_1}$ follows
from the discrete equation (\ref{3}). We can assume 
(and check {\it a posteriori})
that in the late coarsening stage there is a (quasi-steady-state) 
balance between the
processes of attachment and detachment of single atoms by large clusters:
\begin{equation}
- N_1 \sum_{s \geq 2}K_s N_s 
+\sum_{s\geq3}\frac{N_s}{\tau_s} \simeq 0\,,
\label{3a}
\end{equation}
while
the rest of the terms of Eq. (\ref{3}) become irrelevant. Treating $s$ 
as a continuous variable, we obtain
\begin{equation}
N_1 (t) = \frac{\int_0^{\infty} \tau_s^{-1}\, N_s\,ds}
{\int_0^{\infty} K_s\, N_s\, ds}\,.
\label{7}
\end{equation}
(Again, we can 
assume that the number of clusters with small $s$ is very small and
formally shift to zero 
the lower limit of integration over the continuous variable
$s$.) 

The same Eq. (\ref{7}) can be formally obtained if 
we multiply the 
both sides 
of Eq. (\ref{6}) by $s$, integrate over $s$ and use the conservation
law Eq. (\ref{5}). Then, performing
integration by parts in the two remaining terms, we again arrive at Eq.
(\ref{7}). In this derivation one should disregarded 
the boundary terms produced by integration by parts. As 
will be checked later, the ``upper" boundary terms
$s V_s N_s$ and $s (\partial/\partial s) (D_s N_s)$ at $s \to \infty$ vanish. 
In this case the third term vanishes there automatically. 
The boundary terms corresponding to the lower limit 
of integration are assumed to be negligible
compared to the terms that we take into account. Formally, one should
require that $N_s$ vanishes sufficiently fast at $s\to 0$. (Remember, that
the number of single atoms $N_1$ is described separately.)

Eqs. (\ref{6})-(\ref{7}) (supplemented by appropriate initial and
boundary conditions) represent a complete set of equations for 
the late coarsening stage. If one neglects the diffusion term, 
he recovers the Lifshitz-Slyozov-Wagner
description  and corresponding self-similar
asymptotics and scalings for large times \cite{Binder,Gunton}. To make 
this recovery explicit, one should specify
the dependences of $K_s$ and $\tau_s$ on $s$. Looking for
scale invariance, we should assume power laws: $K_s = K_1 s^p$
and $\tau_s = a s^q$. Now, assuming a compact 
cluster morphology ($d$-dimensional spherical
``drop", where $d$ is equal to 2 or 3), I demand 
that Eqs. (\ref{6}) and (\ref{7}) (without the
diffusion term) coincide (after scaling down the coefficients) with
Eqs. (\ref{1}). 
This gives a direct correspondence between
the drift velocities $V_s (t)$, entering
Eq. (\ref{6a}), and $V(R,t)$, 
entering Eq. (\ref{1}). Using this ``correspondence principle" I 
find that the 
concentration of single atoms $N_1 (t)$ scales
like the inverse critical radius $R_c (t)$, while the exponents $p$ and
$q$ must be the following: $p=(d-n-1)/d$ and $q=(n-d+2)/d$.
In particular, for 
the diffusion-controlled
growth ($n=1$) one obtains $p=1/3$ and $q=0$ in three dimensions, and
$p=0$ and $q=1/2$ in two dimensions.
For the interface-controlled growth ($n=0$) one gets $p=2/3$ and
$q=-1/3$ in three dimensions and $p=1/2$ and $q=0$ in two dimensions 
\cite{exponents}. Note that, returning to the original, 
dimensional version of Eq. (\ref{1}) (see, e.g. Refs. \onlinecite{SS}) 
and demanding
an exact coincidence with the zero-diffusion limit of the
cluster model, we can find the coefficients $K_1$ and $a$ as 
well, for every growth mechanism. 

Let us return to the case of a non-zero diffusion in Eq. (\ref{6}). As it is 
seen from the second
formula of Eq. (\ref{6a}), the mapping procedure, described above, completely 
determines the diffusion coefficient $D$. Simple 
scaling 
arguments
show that the diffusion term becomes irrelevant at long 
times \cite{irrelevant}. However, 
even without going into much detail in Eq. (\ref{6}), one can see that
this term
produces (already at $t>0$!) an exponentially small
tail in the DF, even if the DF 
had a compact support at $t=0$. In essence, small fluctuations
transform a strictly bounded DF into an extended one. It was shown 
already by
LS  that an extended DF approaches, for long times,
the {\it limiting} self-similar 
solution \cite{LS}. In the language of Eqs. (\ref{a}) and (\ref{b}),
one can say, therefore, that fluctuations select
\begin{equation}
v_0=\frac{n+2}{n+1} \quad \mbox{and} \quad 
\sigma = \left(\frac{n+2}{n+1}\right)^{n+1}\,.
\label{10}
\end{equation}
Therefore, fluctuations 
provide a strong selection rule
in favor of the limiting solutions of LS (for $n=1$), of Wagner (for $n=0$)
and of its counterparts for other growth mechanisms.

Now we can go back and justify the disregard of the upper
boundary terms arising from
the integration by parts in the derivation of Eq. (\ref{7}). For a
DF with an exponentially small tail at large $s$, the boundary terms obviously
vanish in the scaling regime, as a power-law increase of 
$V_s$ with $s$ is too slow to change anything.
This case is relevant for
an initially
compact DF, as fluctuations produce
an exponentially small tail.
For an extended initial DF with a {\it power-law} tail, $N_s (t=0) \propto
s^{-\mu}$, finiteness of the cluster concentration,
$$
\int_0^{\infty} N_s (t)\, ds < \infty
$$
requires $\mu>2$. Then, using the assumed power laws 
for $K_s$ and $\tau_s$ and evaluating the boundary terms 
$s V_s N_s$ and 
$s (\partial/\partial s) (D_s N_s)$, we must require 
the following inequalities
$p < 1$ and $q > - 1$. Using the 
values of the exponents
$p$ and $q$ determined from the ``correspondence principle", we see
that these two inequalities are satisfied for any $n>-1$, that is for all cases
of physical interest. 

One can suggest the following physical argument 
supporting the strong selection
rule. In 
the 
absence of fluctuations, a bounded
DF always remains bounded \cite{MS96,GMS,Pego}. In other words, 
there is exactly
{\it zero} probability to have clusters with a size larger than some
finite time-dependent size.  On the contrary,
the presence of 
fluctuations leads to a small, but {\it non-zero} 
probability
of the appearance of clusters with 
{\it any} number of atoms. As
the dynamics of OR is very sensitive
to small changes in the region
of the largest available clusters, the presence of the infinite tail in the 
DF will ultimately affect the whole dynamics, driving the DF towards 
the limiting DF.  Of course, as fluctuations 
in macroscopic systems are extremely small,
the time necessary for the DF to actually 
converge to the selected limiting 
solution can be extremely long (if one starts from a bounded
DF). In this case I 
expect that, on a (quite long) intermediate time scale, 
a self-similar DF selected by $\lambda$ will develop, and only 
at much later times
crossover to
the limiting DF will be observed. This crossover can happen much
earlier
if coarsening in {\it mesoscopic} systems is considered, where
the role of discrete nature of atoms increases dramatically. For example, 
I expect this effect to be observable in the processes of
submonolayer relaxation of atomic clusters on surfaces, 
after epitaxial deposition 
is stopped.

Let us compare the cluster approach used in this work 
with the approach of Mullins \cite{Mullins}. Mullins accounted for the
fact that droplets with the same radius do not necessarily have the same
expansion/shrinkage rates (because of correlations). He 
generalized the classical LS-model by replacing the deterministic growth
law for a droplet, $\dot{R} = V (R,t)$ (where $V$ is given by
the second 
equation in (\ref{1}), by the equation $\langle \dot{R}| R\rangle =V(R,t)$,
where  $\langle \dot{R}| R\rangle$ is the {\it average} value of $\dot{R}$
for droplets with a given $R$. Then he inserted
this relationship in the continuity
equation (\ref{1}). In contrast to the cluster approach, the approach of Mullins
does not produce a diffusion term (as it does not account for fluctuations
related to discrete character of particles), therefore it does not provide
a strong selection mechanism for OR.  

Finally, I briefly speculate on possible additional 
mechanisms of strong selection.
An account of fluctuations represents only one of possible ways of going beyond 
the ``classical" LS-formulation. Various finite 
volume fraction effects can provide alternative ways. One such alternative
is rare coagulation events that can be accounted for 
already in the mean-field formulation. This alternative
was briefly discussed by LS already in 1961
\cite{LS}. As the result of the rare coagulation events, a DF that had
a compact support at $t=0$ is also expected to 
develop a tail which will
drive it
towards the limiting solution. No 
quantitative analysis of this scenario is presently 
available. Another possibility
involves correlation effects, completely ignored by any 
mean-field description.
Here I should mention the work of Marder \cite{Marder}
who studied screening effects in OR 
and arrived at a {\it different}
FP equation for the DF. 
In his analysis, the diffusion term results from the screening effects, rather
than from fluctuations, and it is proportional to the square root
of the volume fraction. Of course, the presence of any linear
diffusion term in the FP equation will
produce a tail in the DF and drive the DF towards the limiting
solution
(if the diffusion term is small enough and does not interfere 
in the scaling regime). The comparative role of these possible
selecting mechanisms is obviously volume-fraction-dependent. Of 
course, non-universal transients and
convergence rates towards the selected
DF are expected to differ significantly in the different scenarios.

In summary, small fluctuations 
provide a strong selection rule in the problem
of Ostwald ripening, as they drive the system, at long times, 
towards the limiting 
Lifshitz-Slyozov solution. 

I acknowledge useful discussions with O. Biham and A.J. Vilenkin. 
This work was supported in part by a grant from Israel Science Foundation,
administered by the Israel Academy of Sciences and Humanities.


\begin{references}
\bibitem{irrelevant} Indeed,
the second derivative with respect to $s$
includes an additional small factor of order 
$1/\bar{s}$ that decreases with time. Here $\bar{s}$ is the
characteristic
value of $s$ (for example, the average number of particles in a 
cluster).
\end{references}
\end{document}